\documentclass[utf8]{FrontiersinHarvard} 

\usepackage{url,hyperref,lineno,microtype,subcaption}
\usepackage[onehalfspacing]{setspace}

\def\keyFont{\fontsize{8}{11}\helveticabold }
\def\firstAuthorLast{Landt} 
\def\Authors{Hermine Landt\,$^{1,*}$}
\def\Address{$^{1}$Centre for Extragalactic Astronomy, Department of Physics, Durham University, Durham, UK}
\def\corrAuthor{Centre for Extragalactic Astronomy, Department of Physics, Durham University, South Road, Durham, DH1 3LE, UK}

\def\corrEmail{hermine.landt@durham.ac.uk}

\begin{document}

\def\la{\mathrel{\hbox{\rlap{\hbox{\lower4pt\hbox{$\sim$}}}\hbox{$<$}}}}
\def\ga{\mathrel{\hbox{\rlap{\hbox{\lower4pt\hbox{$\sim$}}}\hbox{$>$}}}}

\def\mnras{MNRAS}
\def\apj{ApJ}
\def\apjl{ApJL}
\def\pasp{PASP}
\def\aap{A\&A}
\def\araa{ARA\&A}

\font\sevenrm=cmr7
\def\OIII{[O~{\sevenrm III}]}
\def\FeII{Fe~{\sevenrm II}}
\def\FeIIf{[Fe~{\sevenrm II}]}
\def\SIII{[S~{\sevenrm III}]}
\def\HeI{He~{\sevenrm I}}
\def\HeII{He~{\sevenrm II}}
\def\NeV{[Ne~{\sevenrm V}]}
\def\OIV{[O~{\sevenrm IV}]}

\def\iraf{{\sevenrm IRAF}}
\def\mpfit{{\sevenrm MPFIT}}
\def\galfit{{\sevenrm GALFIT}}
\def\prepspec{{\sevenrm PrepSpec}}
\def\mapspec{{\sevenrm mapspec}}
\def\cream{{\sevenrm CREAM}}
\def\javelin{{\sevenrm JAVELIN}}
\def\cloudy{{\sevenrm CLOUDY}}
\def\banzai{{\sevenrm BANZAI}}
\def\orac{{\sevenrm ORAC}}
\def\demc{{\sevenrm DEMC}}

\def\gp{\mathcal{GP}}

\onecolumn
\firstpage{1}

\title[Dust in the Disk]{The outer dusty edge of accretion disks in active galactic nuclei} 

\author[\firstAuthorLast ]{\Authors} 
\address{} 
\correspondance{} 

\extraAuth{}

\maketitle

\begin{abstract}

Recent models for the inner structure of active galactic nuclei (AGN) aim at connecting the outer region of the accretion disk with the broad-line region and dusty torus through a radiatively accelerated, dusty outflow. Such an outflow not only requires the outer disk to be dusty and so predicts disk sizes beyond the self-gravity limit but requires the presence of nuclear dust with favourable properties. Here we investigate a large sample of type 1 AGN with near-infrared (near-IR) cross-dispersed spectroscopy with the aim to constrain the astrochemistry, location and geometry of the nuclear hot dust region. Assuming thermal equilibrium for optically thin dust, we derive the luminosity-based dust radius for different grain properties using our measurement of the temperature. We combine our results with independent dust radius measurements from reverberation mapping and interferometry and show that large dust grains that can provide the necessary opacity for the outflow are ubiquitous in AGN. Using our estimates of the dust covering factor, we investigate the dust geometry using the effects of the accretion disk anisotropy. A flared disk-like structure for the hot dust is favoured. Finally, we discuss the implication of our results for the dust radius-luminosity plane.

\tiny
 \keyFont{ \section{Keywords:} Active galactic nuclei --- Quasars --- Dust continuum emission --- Dust physics --- Near-infrared astronomy}
\end{abstract}

\section{Introduction}

A prominent feature in the multi-wavelength spectral energy distributions (SEDs) of active galactic nuclei (AGN) is strong infrared (IR) continuum emission, which originates from thermal dust radiation. The structure of the dust producing it is commonly assumed to be an optically thick toroid, which is aligned with the plane of the accretion disk. The accretion disk then provides the UV/optical photons that heat the dust. An alternative to the dusty torus was already proposed early on by \citet{Phi89b}, namely, an extended and warped dusty disk. Such a structure is attractive since it could present a natural transition between the accretion disk and the central dust, without the stability problem of the torus \citep{Kro07}. 

The extended dust structure emits over a large IR wavelength range, with the near-IR believed to be dominated by the hottest dust located closest to the central supermassive black hole. Therefore, AGN are most suitable to investigate the chemical composition and grain properties of astrophysical dust. Their UV/optical luminosities are usually high enough to heat the central dust to sublimation temperatures, a property which they have in common with protoplanetary disks around young stars. If we can observe these highest temperatures, we can in principle constrain the chemistry since different species condense out of the gas phase in different environmental conditions. Previously dust temperatures were measured in a handful of AGN by obtaining simultaneous photometry at several near-IR wavelengths \citep{Clavel89, Glass04, Schnuelle13, Schnuelle15}, but has only come of age with the availability of efficient near-IR cross-dispersed spectrographs. \citet{L11a} and \citet{L14} derived dust temperatures from such spectroscopy for the largest sample of type~1 AGN so far ($\sim 30$ sources). They found a very narrow dust temperature distribution with an average value of $T \sim 1400$~K. This result, which is similar to what has been found in protoplanetary disks \citep{Monnier05}, either indicates dust at sublimation but composed mainly of silicate grains and so an oxygen-rich environment from which the dust formed. Or, if carbon dominates the composition, then the dust is {\it not} heated to close to sublimation, since carbonaceous dust, e.g. graphite, can survive up to $T \sim 2000$~K \citep{Sal77}.

The first near-IR {\it spectroscopic} monitoring campaign of the hot dust in an AGN, namely, NGC~5548, found that a single component dominated both the mean and variable emission, with the dust reponse time and luminosity-based dust radius being consistent with each other if the emissivity of a pure blackbody was assumed \citep{L19}. Thus, the dust grain size of the hot dust in this AGN could be constrained to relatively large values (of a few $\mu$m). From the estimated dust temperature and its variability they concluded that the dust composition was predominantly carbonaceous and well below the sublimation threshold. The reverberation signal of the dust was then mainly due to a heating and cooling process in response to the variable UV/optical accretion disk flux irradiating it. Most importantly, the dust reverberation signal showed tentative evidence for a second hot dust component most likely associated with the accretion disk. The existence of dust in the outer regions of the accretion disk is a prerequisite for the recent models of the AGN structure proposed by \citet{Czerny17} and \citet{Baskin18}. These authors explain both the broad emission line region (BLR) and the dusty torus as part of the same outflow launched from the outer accretion disk by radiation pressure on dust. Since carbonaceous dust has a higher opacity than silicate dust, the former is preferred in this scenario. Most recently, \citet{L23} presented results from a near-IR spectroscopic monitoring campaign on the high-luminosity AGN Mrk~876. The comparison of the mean and variable (rms) spectrum clearly showed that at least two hot dust components are present in AGN, with the second component possibly originating in the accretion disk. However, contrary to the case of NGC~5548, the independent measure of the dust radius via reverberation mapping yielded a value a factor of $\sim 2$ lower than the luminosity-based dust radius estimate, indicating that the anisotropy effects of the accretion disk illumination and through it the dust geometry can be detected and studied in this way. For Mrk~876, it was concluded that the geometry is most likely a flared, dusty disk with an enlarged 'inner hole', which is very similar to the paradigm commonly assumed for protoplanetary disks.

Given the high potential that a comparison between luminosity-based dust radii and those obtained by independent methods, e.g. by reverberation mapping, carries for revealing the astrochemistry and geometry of the hot dust in AGN, this study sets out to apply it to a large sample of AGN. In Section~\ref{samplesec}, we discuss the selection of the sample, whereas Section~\ref{lumradius} gives details of the measurements required for the estimation of the luminosity-based dust radius. In Section~\ref{discussion}, we discuss our main results and finally present the conclusions in Section~\ref{conclusion}.

\section{The sample selection} \label{samplesec}

As discussed by \citet{L19, L23}, combining the assumption of radiative equilibrium of optically thin dust, which encodes a dependence on the dust chemical species and grain size through the dust emissivity parameter (see eq. \ref{Stefan-Boltz}), with an independent measurement of the dust radius can constrain the dust astrochemistry. Alternative dust radius measurements can come from, e.g., the dust response time estimated by reverberation mapping or the geometric distance of the dust measured through near-IR interferometry. A promising new method is also the estimate of the location of the scattering region through optical spectropolarimetry of the BLR \citep{Shab20}. Therefore, this study required a sample of type 1 AGN with an optical/near-IR continuum dominated by the AGN rather than the host galaxy starlight. Furthermore, the available near-IR spectrum should cover a sufficiently large wavelength range to allow for both an estimate of the accretion disk flux level as well as a measurement of the hot dust temperature in order to be able to calculate the luminosity-based dust radius. This requirement is usually met only by cross-dispersed near-IR spectra \citep{L11a}. Finally, the type 1 AGN with available cross-dispersed near-IR spectra were required to have a published near-IR dust radius based on an alternative method. The total sample comprises of 39 objects and its properties are listed in Table~\ref{sample}. The independent dust radius measurement was mostly from near-IR (photometric) dust reverberation mapping campaigns covering the wavelength region of up to a few $\mu$m and so sampling the SED of the hot dust, and in only 4/39 sources the dust radius measurement was based on near-IR interferometry. 

\begin{table*}
\caption{\label{sample} 
The sample}
\begin{tabular}{lccccrlrl}
\hline
Object name & $z$ & log $L_{\rm uv}$ & $T_{\rm d}$ & log $L_{\rm d}$ & $R_{\rm d,lum}$ & Ref. & $R_{\rm d,rev}$ & Ref. \\
&& (erg s$^{-1}$) & (K) & (erg s$^{-1}$) & (lt-days) & & (lt-days) & \\
(1) & (2) & (3) & (4) & (5) & (6) & (7) & (8) & (9) \\
\hline
3C~351        & 0.372 & 46.63 & 1373 & 46.78 &  792 & R06       & 1203$\pm$542 & Lyu19 \\
PG~2233$+$134 & 0.326 & 46.85 & 1413 & 45.65 &  964 & This work &  343$\pm$29  & Lyu19 \\
PG~0953$+$414 & 0.234 & 46.45 & 1391 & 45.75 &  627 & This work &  566$^{+50}_{-38}$ & M19 \\
PG~0947$+$396 & 0.206 & 45.22 & 1406 & 44.96 &  149 & This work & 1629$\pm$23 & Lyu19 \\
PDS~456       & 0.184 & 47.41 & 1425 & 46.83 & 1806 & L08       & 1599$\pm$213$^\star$ & G20 \\
3C~273        & 0.158 & 47.59 & 1443 & 46.76 & 2166 & L08       & 1656$\pm$59 & Lyu19 \\
PG~0052$+$251 & 0.155 & 46.07 & 1198 & 45.29 &  546 & L13       &  347$\pm$37 & Lyu19 \\
PG~1307$+$085 & 0.155 & 46.22 & 1307 & 45.25 &  545 & L13       &  310$\pm$40 & Lyu19 \\
PG~0026$+$129 & 0.145 & 46.41 & 1127 & 45.15 &  913 & L13       &  487$\pm$36 & Lyu19 \\
PG~1519$+$226 & 0.137 & 45.85 & 1538 & 45.24 &  257 & R06       &  170$\pm$99 & Lyu19 \\
PG~1612$+$261 & 0.131 & 45.67 & 1549 & 45.25 &  206 & R06       &  555$\pm$35 & Lyu19 \\
Mrk~876       & 0.129 & 46.09 & 1306 & 45.55 &  469 & L23       &  334$^{+42}_{-37}$ & M19 \\
PG~1415$+$451 & 0.114 & 45.92 & 1461 & 45.07 &  309 & R06       &  269$\pm$208 & Lyu19 \\
PG~0804$+$761 & 0.100 & 45.97 & 1314 & 45.70 &  405 & L13       &  600$\pm$21 & Lyu19 \\
PG~1211$+$143 & 0.081 & 46.22 & 1337 & 45.10 &  521 & L13       &  338$\pm$83 & Lyu19 \\
Mrk~478       & 0.079 & 46.11 & 1547 & 45.34 &  343 & R06       &  237$\pm$37 & Lyu19 \\
PG~1448$+$273 & 0.065 & 45.31 & 1477 & 44.55 &  150 & R06       &  263$\pm$30 & Lyu19 \\
PG~0844$+$349 & 0.064 & 46.18 & 1190 & 44.76 &  628 & L11       &   99$^{+13}_{-10}$ & M19 \\
Mrk~1513      & 0.063 & 46.33 & 1356 & 45.04 &  575 & L13       &  494$\pm$42 & Lyu19 \\
I~Zw~1        & 0.060 & 45.58 & 1412 & 45.21 &  224 & G12       &  274$\pm$41 & Lyu19 \\
PG~1126$-$041 & 0.060 & 45.65 & 1515 & 45.09 &  211 & R06       &  523$\pm$26 & Lyu19 \\
Mrk~734       & 0.050 & 45.70 & 1597 & 44.48 &  201 & This work &  103$\pm$7  & Lyu19 \\
Mrk~231       & 0.041 & 45.67 & 1529 & 45.71 &  212 & This work &  393$\pm$83$^\star$ & K09 \\
Mrk~841       & 0.036 & 45.22 & 1437 & 44.23 &  143 & P22       &  110$\pm$15 & Lyu19 \\
Mrk~110       & 0.035 & 46.00 & 1452 & 44.56 &  343 & L11       &  117$\pm$6 & K14 \\
Mrk~509       & 0.034 & 45.77 & 1398 & 44.79 &  284 & L08       &  121$\pm$2 & K14 \\
Ark~120       & 0.033 & 45.45 & 1102 & 45.16 &  316 & L08       &  138$\pm$18 & K14 \\
3C~120        & 0.033 & 45.28 & 1389 & 44.59 &  164 & L13       &   94$^{+4}_{-7}$ & R18 \\
Mrk~817       & 0.031 & 45.65 & 1401 & 44.53 &  246 & L11       &   93$\pm$9 & K14 \\
Mrk~290       & 0.030 & 45.36 & 1353 & 44.26 &  189 & L08       &  124$\pm$3 & Lyu19 \\
H~2106$-$099  & 0.027 & 45.28 & 1342 & 44.27 &  175 & L08       &  303$\pm$36$^\star$ & G23 \\  
Mrk~335       & 0.026 & 45.63 & 1308 & 44.40 &  276 & L08       &  168$\pm$6 & K14 \\
Mrk~79        & 0.022 & 44.82 & 1364 & 44.21 &  100 & L11       &   68$\pm$5 & K14 \\
Mrk~1239      & 0.019 & 44.85 & 1443 & 44.64 &   92 & R06       &  189$\pm$30$^\star$ & G23 \\ 
NGC~5548      & 0.017 & 44.50 & 1450 & 44.02 &   61 & L19       &   75$\pm$8 & L19 \\
NGC~7469      & 0.016 & 45.28 & 1551 & 44.19 &  131 & L08       &   85$\pm$1 & K14 \\
NGC~3783      & 0.010 & 45.23 & 1472 & 44.01 &  138 & P22       &  131$^{+25}_{-50}$ & E23 \\
NGC~4593      & 0.009 & 44.72 & 1380 & 43.71 &   87 & L08       &   42$\pm$1 & K14 \\
NGC~4151      & 0.003 & 43.24 & 1328 & 43.06 &   17 & L08       &   46$\pm$1 & K14 \\
\hline
\end{tabular} 

\parbox[]{16.5cm}{The columns are: (1) object name; (2) redshift; (3) total accretion disk luminosity; for a blackbody emissivity (4) dust temperature; (5) total dust luminosity and (6) dust radius; (7) reference for the cross-dispersed near-IR spectral data used for the continuum fits; (8) near-IR dust reverberation lag time in the rest-frame taken from reference in (9). References are E23: Esser et al. (2023), G12: \citet{GarciaR12}, G20: \citet{gravity20}, G23: \citet{gravity23}, K09: \citet{Kish09}, K14: \citet{Kosh14}, L08: \citet{L08a}, L11: \citet{L11a}, L13: \citet{L13}, L19: \citet{L19}, L23: \citet{L23}, Lyu19: \citet{Lyu19}, M19: \citet{Min19}, P22: \citet{Prieto22}, R06: \citet{Rif06} and R18: \citet{Ram18}.}

\parbox[]{16.5cm}{$^\star$ near-IR interferometric dust radius}

\end{table*}

\section{The luminosity-based dust radius} \label{lumradius}

The calculation of the luminosity-based dust radius requires a measurement of the dust temperature and an estimate of the UV/optical (accretion disk) luminosity that heats the dust. Cross-dispersed near-IR spectra with their relatively large wavelength range cover about half the hot dust SED in low-redshift AGN and additionally a considerable part of the accretion disk spectrum. The latter is expected to be the dominant contributor to the continuum flux up to $\sim 1~\mu$m \citep{L11b, L11a}. In order to obtain the necessary measurements, we decomposed the spectral continuum into these two components, accretion disk and dust emission, following the approach described in \citet{L19}. As already noted in \citet{L19}, due to the relatively small spectral aperture usually used for near-IR spectroscopy, the contribution of host galaxy light to the total observed continuum flux is expected to be negligible in most AGN. In short, we have first approximated the rest-frame wavelength range of $\la 1~\mu$m with the spectrum of a standard accretion disk, which we have subsequently subtracted from the total spectrum. For the calculation of the accretion disk spectrum we adopted the black hole mass listed in \citet{Bentz15} for a geometrical scaling factor of $f=4.3$, where available, and else we have estimated it based on the relationship between black hole mass and near-IR virial product presented in \citet{L13}. The scaling of the accretion disk spectrum to the near-IR spectrum then readily gives the accretion rate. Furthermore, we assumed that the disk is relatively large and extends out to $r_{\rm out}=10^4 r_{\rm g}$. We then fitted the resultant hot dust spectrum at wavelengths $>1~\mu$m with a blackbody, representing emission by large dust grains. Table~\ref{sample} lists the physical parameters derived from the near-IR spectral decomposition. As in \citet{L19}, we then calculated luminosity-based dust radii, $R_{\rm d,lum}$, from the best-fit dust temperatures assuming radiative equilibrium between the luminosity of the irradiating source and the dust:

\begin{equation}
\label{Stefan-Boltz}
\frac{L_{\rm uv}}{4 \pi R_{\rm d,lum}^2} = 4 \sigma T^4 \langle Q^{\rm em} \rangle,
\end{equation}

\noindent
where $\sigma$ is the Stefan-Boltzmann constant and $\langle Q^{\rm em} \rangle$ is the Planck-averaged emission efficiency. We have approximated $L_{\rm uv}$ with the accretion disk luminosity, as given in Table~\ref{sample}, and have used $\langle Q^{\rm em} \rangle = 1$, which is appropriate for the case of a blackbody. The blackbody case is reached for grain sizes of $a \ga 0.4~\mu$m and $\ga 2~\mu$m for carbon and silicate dust, respectively \citep[see Fig. 8 of][]{L19}. As \citet{L19, L23} have shown, radii for dust composed of small grains ($a\la0.1~\mu$m) are a factor of $\sim 6$ and $\sim 10$ larger if the composition is mainly carbonaceous or silicates, respectively. The error on the luminosity-based dust radius is determined by the error on the temperature and the error on the total accretion disk luminosity. Due to the relatively large wavelength coverage of the cross-dispersed near-IR spectra, which are also of relatively high signal-to-noise, the error on the temperature is small \citep[$\sim 10-30$~K;][]{L19, L23}. The accretion disk luminosity is mostly determined by the accretion rate, which is set by the flux level of the near-IR spectrum. As discussed in \citet{L19, L23}, since the telluric standard star is observed close in time with the science target, photometric correction factors are on average $\sim 10-15\%$. Therefore, errors on the luminosity-based dust radius are assumed to be $\sim 10\%$.

\section{Results and discussion} \label{discussion}

Based on two well-studied sources, namely, NGC~5548 and Mrk~876, \citet{L23} have recently suggested that there are considerable similarities between the hot dust in AGN and that present in protoplanetary disks around young stars. In particular, these similarities are: (i) the prevalance of large dust grain sizes, which in protoplanetary disks might have formed in the midplane of the disk by differential settling; (ii) the presence of an 'inner hole' beyond the dust-free zone expected due to dust sublimation, which in protoplanetary disks is thought to be a cavity filled with gaseous disk material that can change its optical thickness and thus influence the location of the puffed-up so-called 'wall' or 'inner rim'; and (iii) a general temperature-radius relationship consistent with that for a dusty, flared and passively illuminated protoplanetary disk. Here we will test and expand on this proposition by using a sample of $\sim 40$~AGN.

\subsection{The prevalence of large grains in the hot dust of AGN}

\begin{figure}
\begin{center}
\includegraphics[width=10cm]{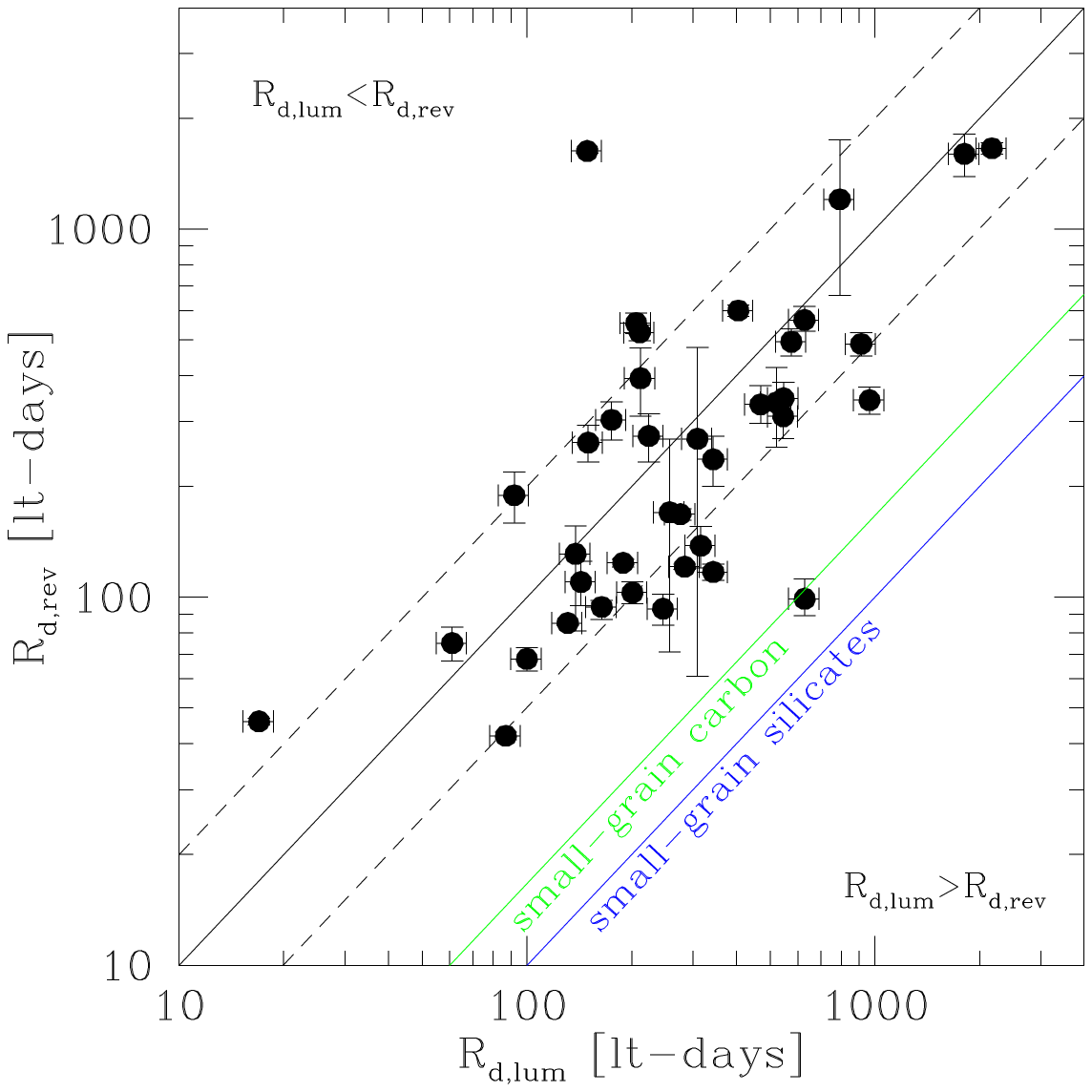}
\end{center}
\caption{\label{radius} The logarithmic near-IR dust reverberation lag time versus the logarithmic luminosity-based dust radius for large dust grain sizes. The solid black line indicates the line of equality, which corresponds to the mean ratio between the two radii, whereas the dashed black lines mark the $1~\sigma$ region around it, corresponding to a factor of 2. Luminosity-based dust radii corresponding to small-grain ($a\la0.1~\mu$m) carbon and silicate dust are expected to be larger than the values for large dust grains by a factor of $\sim 6$ (solid green line) and $\sim 10$ (solid blue line), respectively.}
\end{figure}

As discussed in \citet{L23}, the assumption of dust, which is optically thick to the UV/optical radiation heating it and optically thin to its own IR radiation, allows us to constrain the astrochemistry of the dust through the dust emissivity parameter $Q^{\rm em}$ (see eq. \ref{Stefan-Boltz}), {\it if the dust radius can be measured by an independent method.} Fortunately, several such methods exist, including, e.g., the dust response time measured through reverberation mapping and a geometric dust radius measurement based on near-IR interferometry. Here we have compiled dust radius measurements from these two independent methods for a sample of AGN with available near-IR spectra that were suitable to derive luminosity-based dust radii. Fig.~\ref{radius} plots the comparison between the luminosity-based dust radius and that obtained by an independent method, which was mostly near-IR photometric reverberation mapping. We note that the two dust radii are generally not contemporaneous and so could be affected by the variability effects of the irradiating luminosity, although \citet{L19} argued that the hot dust radius is not set by sublimation and is probably luminosity-invariant.  

Fig.~\ref{radius} shows that the large majority of the data points cluster around the line of equality for a blackbody emissivity, which corresponds to the largest dust grains. The mean ratio is $R_{\rm d,lum}/R_{\rm rev} = 1$, with a dispersion around the mean of a factor of $\sim 2$. We note that for most sources the luminosity-based dust radius is {\it larger} than the dust radius measured by the independent method. Assuming instead an emissivity corresponding to small dust grains of predominantly carbonaceous or silicate composition would move the line of equality to luminosity-based dust radii larger by a factor of $\sim 6$ (green solid line) and $\sim 10$ (blue solid line), respectively. Although in a handful of sources this possibility cannot be excluded, in general small grains do not seem to dominate the hot dust composition.

\subsection{A flared, passively illuminated (hot) dusty disk with an 'inner hole' in AGN}

\begin{figure}
\begin{center}
\includegraphics[width=8.5cm]{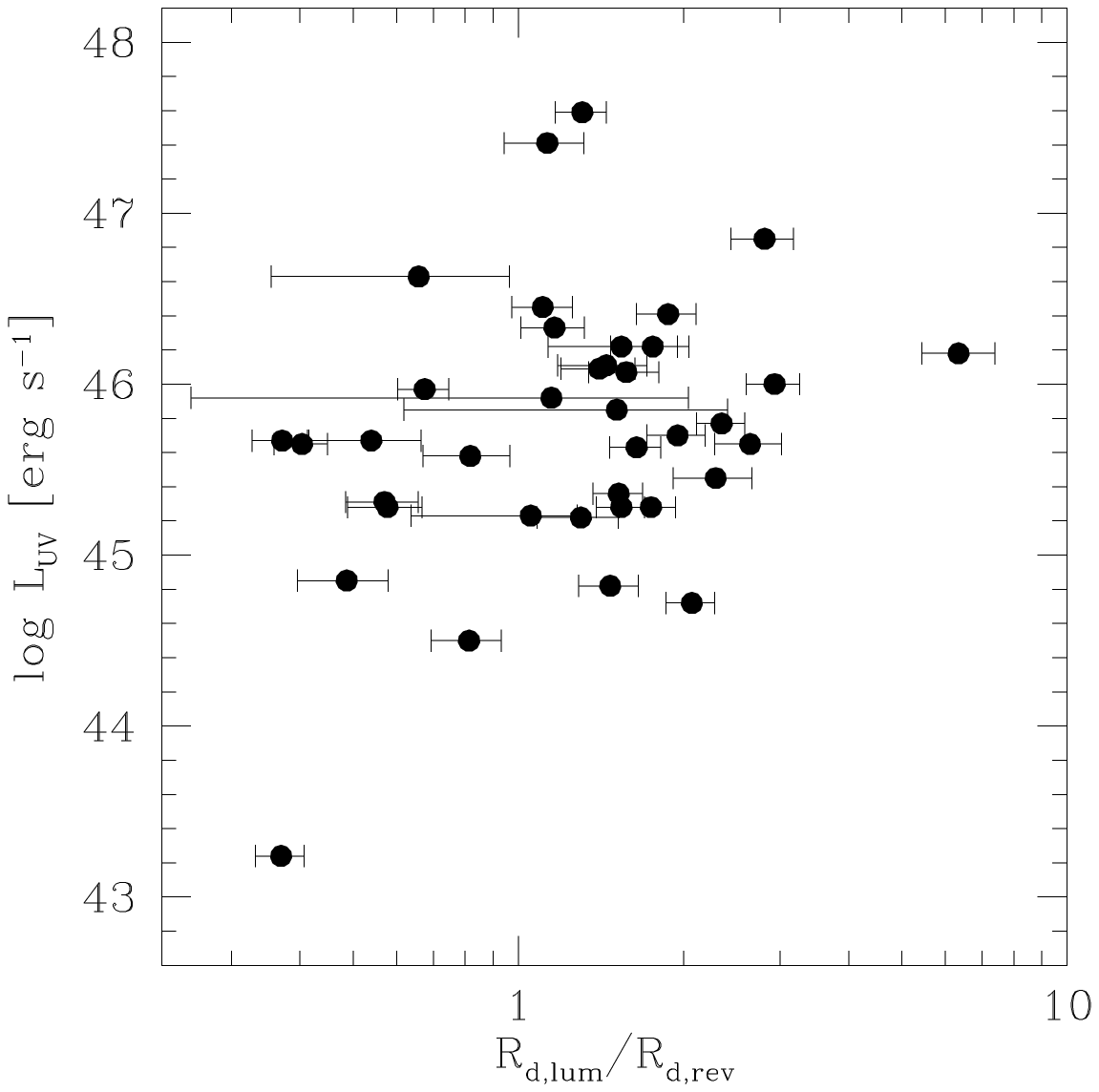}
\includegraphics[width=8.5cm]{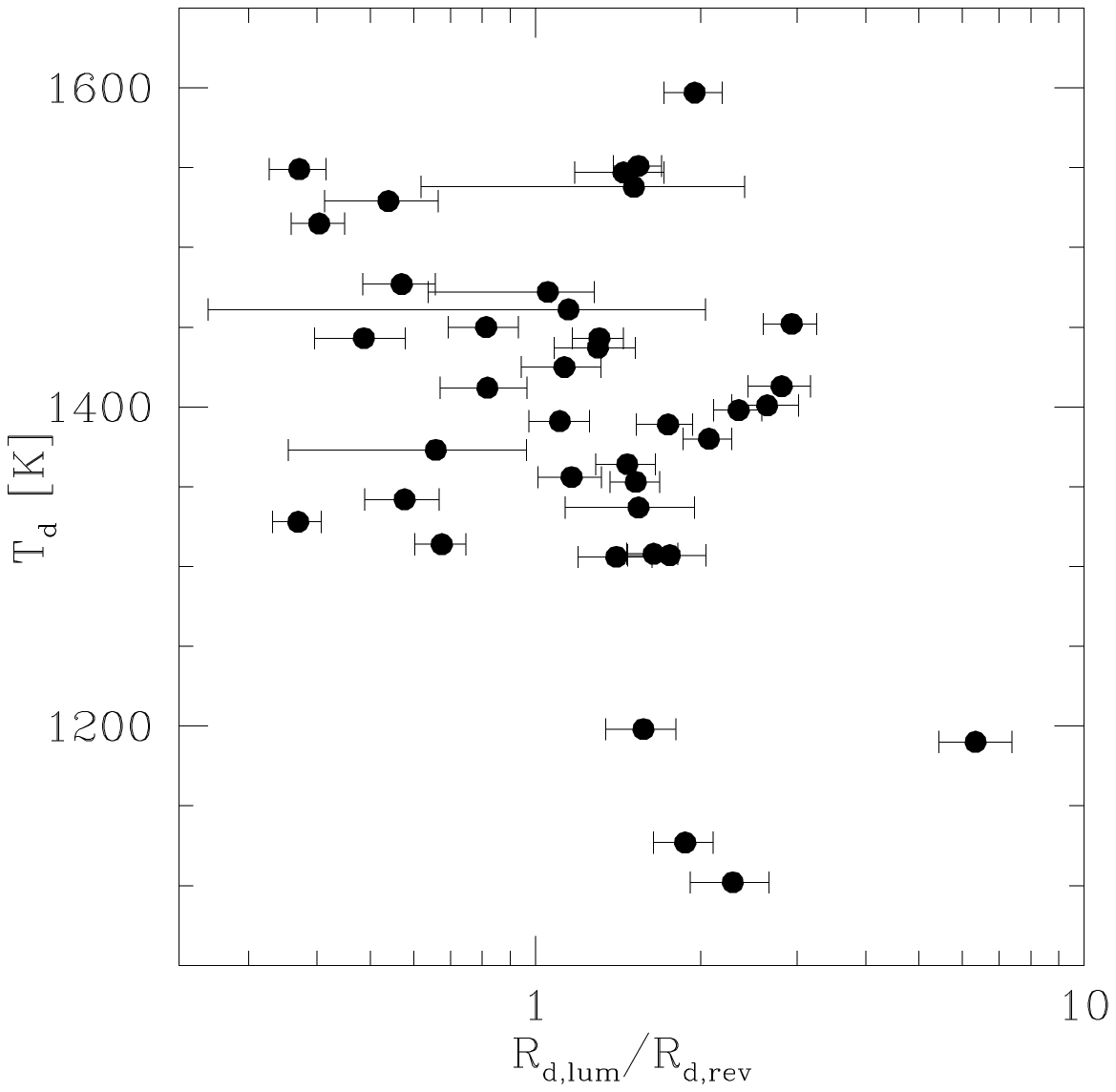}
\end{center}
\caption{\label{lbol} The logarithmic total accretion disk luminosity (left panel) and the dust temperature (right panel) versus the logarithm of the ratio between the luminosity-based dust radius and the dust radius measured by reverberation mapping (or near-IR interferometry).}
\end{figure}

It is in general of high interest to understand the origin of the scatter in Fig.~\ref{radius} and if it is caused by AGN physics rather than being mainly due to variability and/or measurement uncertainties. Such an investigation is particularly timely, since considerable research is under way to help establish the hot dust radius as a suitable standard candle for cosmological studies using AGN \citep[e.g.][]{Hoenig17}. A first indication that a considerable inconsistency can be present between the luminosity-based hot dust radius and that obtained by other means was found for the high-luminosity AGN Mrk~876 by \citep{L23}. The difference between the two values of a factor of $\sim 2$ could be well explained if the dust was assembled in a flared, disk-like geometry, which is naturally expected to be carved out by the anisotropy of the accretion disk illumination \citep{Kaw10, Kaw11}. Then, the dust is expected to be illuminated by the UV/optical accretion disk luminosity reduced by a factor $\cos \theta$, with $\theta$ the angle between the accretion disk rotation axis and the location of the dust. Since such a pronounced difference was not found for the low-luminosity AGN NGC~5548 by the study of \citet{L19}, there is a strong possibility that the anisotropy effect depends on AGN luminosity. An indication that the accretion disk illumination anisotropy considered by \citet{Kaw10, Kaw11} affects low- and high-luminosity AGN differently was also found by \citet{Min19}. Their sample showed a best-fit slope for the logarithmic reverberation dust radius versus optical luminosity relationship of $\sim 0.4$, i.e. shallower than the slope of 0.5 predicted by radiative equilibrium considerations (see eq.~\ref{Stefan-Boltz}).

Fig.~\ref{lbol} (left panel) investigates this conjecture, where we plot the total accretion disk luminosity estimated from the fits to the near-IR spectroscopy versus the ratio between the luminosity-based dust radius and the dust radius measured by reverberation mapping (or near-IR interferometry). There is clearly a trend for the dust radius ratio to increase with luminosity and so for the disk structure to be more pronounced in high-luminosity sources. The significance of a linear correlation is $P=95.5\%$. Fig.~\ref{lbol} (right panel) plots the dependence of the dust radius ratio on the dust temperature. The dust temperatures reach values well above the sublimation temperature for silicates (of $\sim 1400$~K) and, therefore, the hot dust in AGN is most likely carbonaceous, as already argued previously \citep[e.g.][]{Mor09, L11b, L19}. A trend is present also in Fig.~\ref{lbol} (right panel) that indicates that the 'inner hole' of the disk is more enlarged in high-luminosity sources, if we assume a single chemical species for the hot dust and so a unique sublimation temperature (e.g. $\sim 1900$~K, corresponding to carbonaceous dust). The significance of a linear correlation is $P=93.5\%$. However, we note that the scatter is considerable in both relationships displayed in Fig.~\ref{lbol}.

\subsection{The outer dusty edge of accretion disks in AGN}

\begin{figure}
\begin{center}
\includegraphics[width=10cm]{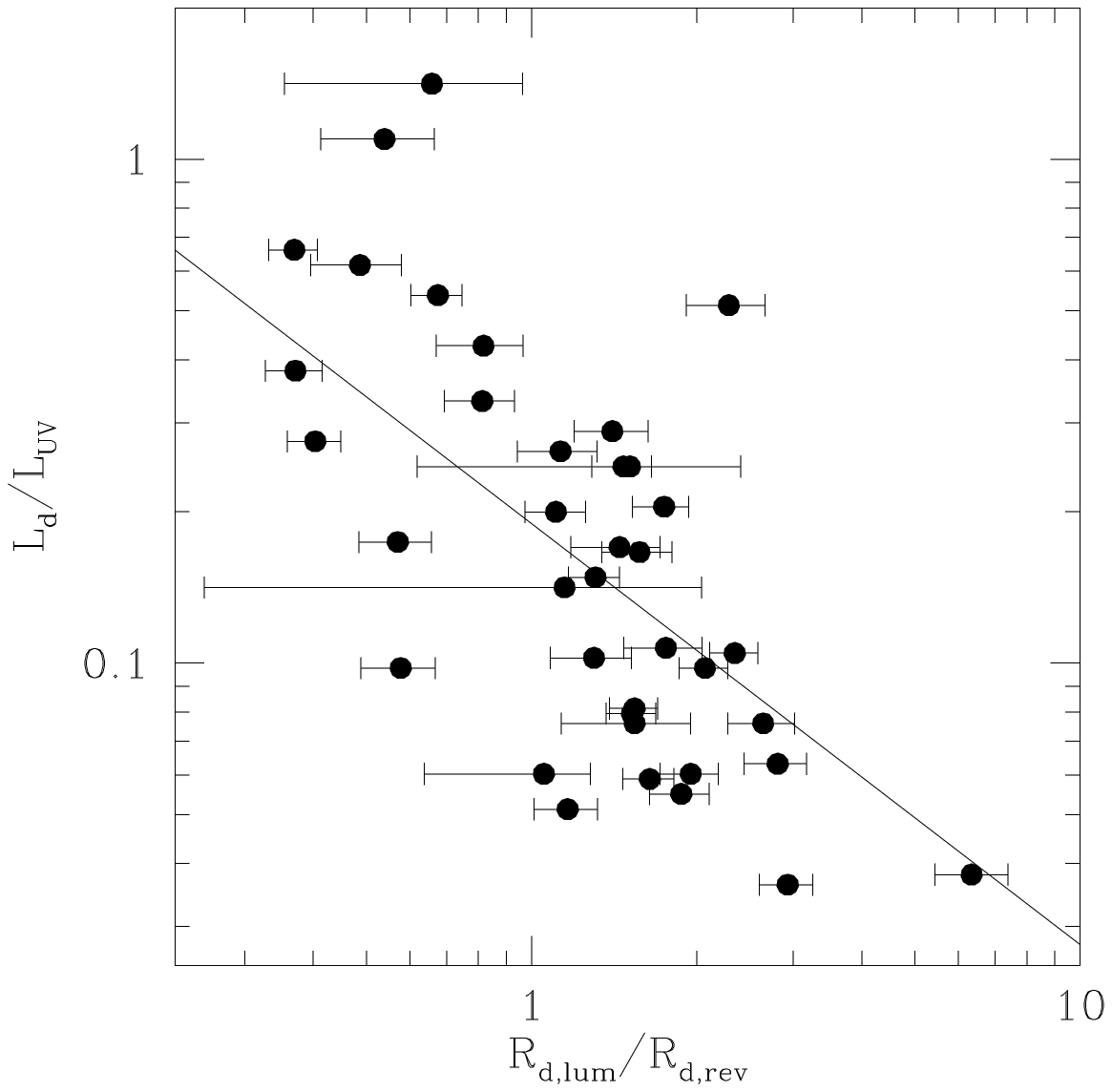}
\end{center}
\caption{\label{covfac} The logarithmic hot dust covering factor versus the logarithm of the ratio between the luminosity-based dust radius and the dust radius measured by reverberation mapping (or near-IR interferometry). The best-fit relationship is shown as the black solid line. See text for more details.}
\end{figure}

\begin{figure}
\begin{center}
\includegraphics[width=10cm]{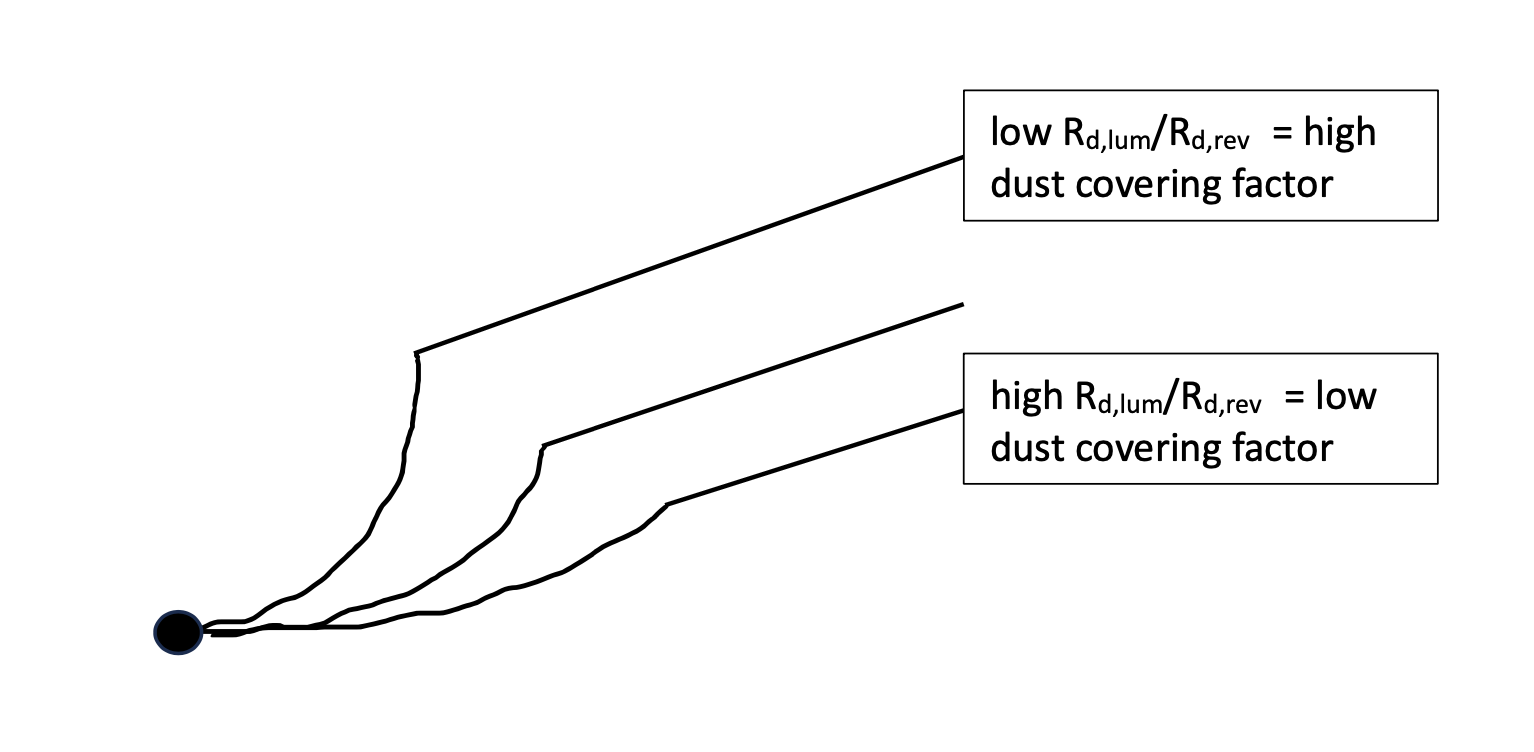}
\end{center}
\caption{\label{model} Sketch of a possible change in the dust geometry as the ratio between the luminosity-based dust radius and the dust radius measured by reverberation mapping (or near-IR interferometry) increases and thus the dust covering factor decreases.}
\end{figure}

A considerably tighter relationship is present between the hot dust covering factor, defined as the ratio between the total dust luminosity and the total accretion disk luminosity, and the dust radius ratio (see Fig.~\ref{covfac}). The significance of a linear correlation is $P>99.999\%$ for a best-fit relationship of $\log (L_{\rm d}/L_{\rm UV}) = (-0.72\pm0.05) - (0.84\pm0.15) \cdot \log (R_{\rm d,lum}/R_{\rm d,rev})$. This relationship could be interpreted as evidence that the ratio between the luminosity-based dust radius and that measured by reverberation mapping (or near-IR interferometry) is a suitable indicator of the dust geometry. Then, the larger this ratio, the more pronounced the flared, dusty disk structure and so the lower the expected dust covering factor. In other words, as the ratio increases, the dust geometry moves from almost circular to a short disky component with a large flare and eventually to a long disky component with a smaller flare. A sketch of this possible change in dust geometry is exemplified in Fig.~\ref{model}. \citet{Baskin18} investigated the geometry of a dusty, inflated accretion disk structure. Whereas they assumed largely varying grain sizes (spanning about two orders of magnitude), which resulted in the sublimation process occuring over a transition zone rather than at a single location, the scale height of this geometry could be related to the change in dust geometry proposed here. In their model, the disk height is set by the balance between the effects of gravity and radiation pressure and depends only on the accretion rate and the opacity of the material (see their eq. 22). Therefore, it predicts that more luminous AGN have higher dust covering factors, which is contrary to the observed relationship presented in Fig.~\ref{lbol}.   

Recently, \citet{L23} presented the comparison of the mean and variable (rms) spectrum in the high-luminosity AGN Mrk~876, which showed clearly that at least two hot dust components are present in AGN. They found that the total dust emission was dominated (at the $\sim 70\%$~level) by the non-variable component, which could have its origin in the outer (flared) part of the dusty disk or the outer dusty edge of the accretion disk or both. How would such a second hot dust component manifest itself in the relationship shown in Fig.~\ref{covfac}? We would need to know the viewing angle, $\theta$, in order to estimate the accretion disk luminosity as seen by the dust. Then, the true dust covering factor should be calculated relative to this (reduced) $L_{\rm UV}$. In other words, if the main origin of the second dust component is from the flare, the more correct our already assumed accretion disk luminosity is. However, if the main origin of the second dust component is instead in the accretion disk, we should expect to have {\it overestimated} the accretion disk luminosity as seen by the dust and, therefore, {\it underestimated} the dust covering factor. If indeed the dust radius ratio is a suitable indicator of the dust geometry as described above, as this value increases, the flare becomes less important and the accretion disk dust more important, thus apparently reducing the dust covering factor and tightening the relationship displayed in Fig.~\ref{covfac}. We note that although the dust in the accretion disk will not necessarily be externally illuminated, if it is composed of $\mu$m-sized grains, it will be well-coupled to the accretion disk gas and so affected by its (transmitted) total luminosity in a similar way. \citet{Sta16} investigated in detail the effects of the anisotropy of the accretion disk luminosity on the covering factor for the torus dust. They found that for type~1 AGN low and high covering factors will be underestimated and overestimated, respectively, with the effect being more pronounced for the former than for the latter (see their Fig. 10). Their results could partly explain the steep slope we find for the relationship in Fig.~\ref{covfac} and also the fact that we find dust covering factors exceed unity as well as extremely low values (of only a few per cent).

\section{Conclusions} \label{conclusion}

This study used a sample of $\sim 40$~AGN with available cross-dispersed near-IR spectroscopy to perform a comparison between luminosity-based dust radii and those obtained by independent methods, e.g. by reverberation mapping. The main aim was to reveal the astrochemistry and geometry of the hot dust in AGN. The main results can be summarized as follows.

(i) It could be shown that luminosity-based dust radii obtained with the assumption of a blackbody emissivity were consistent within a factor of $\sim 2$ with dust radii obtained by independent methods. Therefore, large dust grains (with sizes of a few $\mu$m) appear to be ubiquitous in the hot dust of AGN. Assuming instead an emissivity corresponding to small dust grains implied much larger luminosity-based dust radii (by factors of $\sim 6-10$).

(ii) We found a tight relationship between the dust covering factor and the ratio between the luminosity-based dust radius and the dust radius measured by reverberation mapping {\bf (or near-IR interferometry)}. This new relationship can be understood if one assumes that the dust radius ratio is a suitable indicator of the dust geometry and that hot dust is present in the accretion disk in addition to that assembled in a flared, dusty disk, as proposed by \citet{L23}. 

(iii) The effects of the anisotropy of the accretion disk illumination might be different for low- and high-luminosity AGN, as suggested by the flattening of the slope at high luminosities in the dust radius-luminosity plane of \citet{Min19}. We find a trend that the dust radius ratio increases with AGN luminosity, indicating a change in dust geometry whereby the disk structure becomes more pronounced in high-luminosity AGN.

\section*{Conflict of Interest Statement}
The author declares that the research was conducted in the absence of any commercial or financial relationships that could be construed as a potential conflict of interest.

\section*{Author Contributions}
The first author has performed all original measurements and analysis of the data, which were not already reported elsewhere. The first author has written the entire manuscript.

\section*{Funding}
HL acknowledges a Daphne Jackson Fellowship sponsored by the Science and Technology Facilities Council (STFC), UK, and support from STFC grants ST/P000541/1, ST/T000244/1 and ST/X001075/1.



\section*{Data Availability Statement}
The near-IR spectra used for the first time in this work are available on request from the first author.

\bibliographystyle{Frontiers-Harvard} 

\bibliography{references}





\end{document}